\documentclass[apj]{emulateapj}
\usepackage{apjfonts} \usepackage{amsmath} \usepackage{lipsum}
\usepackage{ccaption}
\newcommand{\me}{$M_{\oplus}$} \newcommand{\re}{$R_{\oplus}$}
 \newcommand{\teff}{$T_{\rm eff}$}
\newcommand{\tint}{$T_{\rm int}$} 
\newcommand{\cp}{\citep} \newcommand{\ct}{\citet}
\newcommand{\MNep}{M_{\rm Nep}}

\shorttitle{Planet Nine Characteristics and Detection}
\shortauthors{Fortney et al.}

\begin{document} \title{The Hunt for Planet Nine: Atmosphere, Spectra,
Evolution, and Detectability}

\author{Jonathan J. Fortney\altaffilmark{1}, Mark S.
Marley\altaffilmark{2}, Gregory Laughlin\altaffilmark{1}, Nadine
Nettelmann\altaffilmark{3}, Caroline V. Morley\altaffilmark{1}, Roxana
E. Lupu\altaffilmark{2,4}, Channon Visscher\altaffilmark{5}, Pavle
Jeremic\altaffilmark{1}, Wade G. Khadder\altaffilmark{1}, Mason
Hargrave\altaffilmark{1}}

\altaffiltext{1}{Department of Astronomy \& Astrophysics, University of
California, Santa Cruz, jfortney@ucsc.edu} \altaffiltext{2}{NASA Ames
Research Center, Mountain View, CA, United States}
\altaffiltext{3}{University of Rostock} \altaffiltext{4}{Bay Area
Environmental Research Institute} \altaffiltext{5}{Dordt College}

\begin{abstract} We investigate the physical characteristics of the
Solar System's proposed Planet Nine using modeling tools with a 
heritage in studying Uranus and Neptune.  For a range of plausible
masses and interior structures, we find upper limits on the intrinsic
\teff, from $\sim$35-50 K for masses of 5-20 \me, and we also explore lower \teff\ values.  Possible planetary
radii could readily span from 2.7 to 6 \re\ depending on the mass fraction
of any H/He envelope.  Given its cold atmospheric temperatures, the planet encounters
significant methane condensation, which dramatically alters the
atmosphere away from simple Neptune-like expectations.  We find the
atmosphere is strongly depleted in molecular absorption at visible
wavelengths, suggesting a Rayleigh scattering atmosphere with a high
geometric albedo approaching 0.75.  We highlight two diagnostics for the
atmosphere's temperature structure, the first being the value of the
methane mixing ratio above the methane cloud.  The second is the
wavelength at which cloud scattering can be seen, which yields the
cloud-top pressure.  Surface reflection may be seen if the atmosphere is
thin.  Due to collision-induced opacity of H$_2$ in the infrared, the
planet would be extremely blue (instead of red) in the shortest
wavelength WISE colors if methane is depleted, and would, in some cases,
exist on the verge of detectability by WISE.  For a range of models,
thermal fluxes from $\sim3-5$ $\mu$m are $\sim$~20 orders of magnitude
larger than blackbody expectations.  We report a search of the AllWISE
Source Catalog for Planet Nine, but find no detection.

\end{abstract}

\maketitle

\section{Introduction}\label{introduction}

Recently, \ct{BatyginBrown2016} have found that the orbits of long
period Kuiper Belt Objects cluster in their arguments of perihelion and
also in physical space in the solar system.  They suggest that dynamical
perturbations arising from a relatively small giant planet of perhaps 10
\me, on an eccentric orbit with a semi-major axis distance of perhaps $a
\sim$~700 AU, can explain this clustering.

The formation, evolution, and detectability of this possible ``Planet
Nine'' has been the target of intense study
\cp{Marcos16,Malholtra16,Li16,Linder16,Cowan16,Fienga2016,Ginzburg2016,
BrownBatygin2016,Kenyon16,Bromley16,Mustill16,Holman16}.  To date, studies that have
included discussion of the atmosphere, emission, or cooling history of
the planet have focused on Planet Nine as a blackbody.  This is
certainly a valid starting point, especially since nothing is yet known
about this \emph{possible} object.  Yet a straightforward, but more
detailed look at the planet yields a better understanding of the
planet's atmosphere, evolution, and spectra that cannot come out of
simple analysis.  For instance, the spectra of the solar system's giant
planets differ strongly from blackbodies.

Planet Nine, if it exists, may have a mass between that of Earth and
Uranus. Exoplanets in this mass range are extremely common
\cp{Borucki11}. Should Planet Nine be detected, the characterization of
the planet would either confirm or refute studies such as the one we
present here, and could serve as a proxy for the physical and
atmospheric properties of the as-yet largely unstudied worlds that are
among the most common outcomes of the planetary formation process.

In this Letter we first investigate the cooling history of the Planet
Nine, over a wide range of possible masses from 5--50 \me.  Here we assume a planet
that has at least some H/He atmosphere, in analogy with Uranus and
Neptune as well as the thousands of sub-Neptune exoplanets that suggest
a non-negligible H/He envelope \cp{Lopez14}.  We draw connections
between the cooling history of Planet Nine and our incomplete
understanding of the thermal evolution of Uranus and Neptune.  With
fluxes from the planet's interior and from the Sun then reasonably
well-bounded, we can begin to estimate the atmospheric temperature
structure.  We assess the condensation of methane, atmospheric opacity
sources, and the spectrum of the planet from optical to thermal infrared
wavelengths, focusing on novel and at times counter-intuitive behavior
at very low temperature ($\sim 30-40$ K).  Finally, we apply these
findings to a revised assessment of the detectability of Planet Nine and
report on a new search of the WISE catalog.

\section{Thermal Evolution}

\subsection{Method}

In order to estimate the possible radius and intrinsic luminosity of
Planet Nine we have run Neptune-like thermal evolution models, where we
varied several structure parameters around those appropriate for
Neptune.  These models are somewhat similar to those of \ct{Linder16},
although we make different choices about particular compositions to
investigate. Here we assume planet masses between 5 and 50 \me\
($\MNep=17.15$ \me) and two compositionally different layers consisting
of a H/He envelope atop a core that is either ice-rich (2$\,:\,$1 ice$\,:\,$rock) or ice-poor (1$\,:\,$4 ice$\,:\,$rock).  We
varied the core mass fraction within 0.1--0.3 for the most massive
models, representing a sub-Saturn mass gas giant, 0.30--0.84 for a
15\me\ series of models, and 0.6--0.9 for the least massive models of
5--8 \me, representing a sub-Neptune.  For reference, constraints from
Neptune suggest $\sim$ 0.76--0.90 ice/rock by mass
\citep{Helled11,Nettelmann13}, and detailed modeling suggests a
ice-enriched H/He envelope atop very ice/rock rich deep interior
\citep{Hubbard95,Helled11,Nettelmann13}.  Like \citet{Nettelmann13} we
represent the warm fluid ices in the interior by the water equation of
state (EOS) H$_2$O-REOS.1, H/He the updated H/He-REOS.3 tables of
\citet{Becker14}, and for the rock EOS \citet{HubbardMarley89}.

We followed the thermal evolution of Planet Nine at 1000 AU,
corresponding to a zero-albedo equilibrium temperature of 9 K.  We
explored a range of incident fluxes but found that is yielded little
impact on the thermal evolution.  For the upper boundary condition, the
radiative atmosphere, we applied an analytic description
\citep{Guillot95} of to the  non-gray \ct{Graboske75} model atmosphere
grid, which relates the 1-bar temperature (and hence the specific
entropy of the interior adiabat) and surface gravity to the \teff.  For
the constant of proportionality between the 1-bar temperature and the
\teff\ we use $K=1.50$ based on Neptune.  We assume a hot start initial
condition \cp[e.g.][]{Marley07} and then adiabatic cooling thereafter. 
At all points in the evolutionary history we track the evolution of the
intrinsic luminosity and radius.  See \citet{Nettelmann13} for further
modeling details.

\subsection{Results}

Our modelling results for Planet Nine are summarized in Table 1.  We
recover the strong influence of the assumed H/He mass fraction on the
resulting radius, as it is well known from mass-radius relations for
warm exoplanets \citep[e.g.,][]{Fortney07,Lopez14}. In particular, the
radius of Planet Nine may reach from 2.7 to 4.4 \re\ if its mass is
within 5--8 \me, or from 3.6 to 6.7 \re\ if 15--20 \me\ (Neptune: 3.9
\re), and further to 6.4--8.3 \re\ if 35--50 \me. Intrinsic temperatures
range from the Uranian upper limit of $T_{\rm int} \sim 40 K$ if $M \leq
12$\me to sub-Saturn like values of $T_{\rm int}\sim 60$  K for $M =50$
\me. As expected, we find $T_{\rm int}$ to rise with planet mass, but,
surprisingly, to decrease with H/He content.  The latter behavior
results from the larger energy that can be radiated away from a larger
surface area, and from the low internal temperatures along the H/He
adiabats at $\sim 1$ Mbar pressures.

On the whole, we find excellent agreement with the findings of \ct{Linder16}.  We agree that the planet's luminosity is likely strongly dominated by intrinsic flux, rather than re-radiated solar energy.  Specifically, they suggest a \teff\ of 47 K for a 10 \me\ planet model with a 1:1 ice:rock core and 1.4 \me\ envelope, which suggests an agreement within a few Kelvin for our cooling models, based on our Table 1.

These results should be taken with significant caution.  While adiabatic
cooling models can reproduce the current intrinsic flux from Neptune,
they significantly overestimate the flux from within Uranus, which has
been an outstanding problem for decades \cp{Hubbard95,Fortney10}. 
Therefore these \tint\ and \teff\ values should be taken as upper
limits.  Below, when modeling the atmosphere we examine \tint\ values down to 20 K, lower than that predicted by our evolutionary calculations.

\begin{deluxetable}{cccccccc}[b!] \tabletypesize{\scriptsize}
\tablecolumns{7} \tablewidth{0pt} \tablehead{ \colhead{Planet} &
\colhead{Core} & \colhead{H/He Env.} & \colhead{I:R} & \colhead{Planet} &
\colhead{$T_{\rm eff}$} & \colhead{$T_{\rm 1bar}$} & \colhead{$T_{\rm
int}$} \\ \colhead{Mass (\me)} & \colhead{Mass} & \colhead{Mass} &
\colhead{Ratio} & \colhead{Radius (\re)} & \colhead{(K)} & \colhead{(K)} & \colhead{(K)}
}
\startdata
 5  & 3   & 2   & 2:1 & 4.12  & 36.7  & 51.6  & 36.5  \\
 5  & 4.5 & 0.5 & 2:1 & 2.94 & 42.2  & 54.6  & 41.7   \\
 5  & 4.5 & 0.5 & 1:4 & 2.71 & 38.9  & 48.1  & 36.8   \\
 8  & 5   & 3   & 2:1 & 4.44  & 39.5  & 55.4  & 39.2   \\
 8  & 7   & 1   & 2:1 & 3.44  & 46.2  & 59.7  & 47.2   \\
  8  & 7   & 1   & 1:4 & 3.17  & 42.6  & 52.5  & 40.8   \\
 10 & 5   & 5   & 2:1 & 5.09  & 40.3  & 55.3  & 40.1  \\ 
 10 & 9   & 1   & 2:1 & 3.46  & 48.3  & 60.9  & 47.8    \\ 
 10 & 9   & 1   & 1:4 & 3.16  & 45.1  & 54.3  & 42.2    \\  
 12 & 5   & 7   & 2:1 & 5.57  & 41.4  & 57.1  & 41.2  \\
 12 & 10  & 2   & 2:1 & 3.93  & 46.1  & 58.2  & 45.7  \\ 
 12 & 10  & 2   & 1:4 & 3.64  & 44.6  & 52.3  & 42.8  \\  
 15 & 5   & 10  & 2:1 & 6.16  & 43.2  & 62.4  & 42.8  \\ 
 15 & 13  & 2   & 2:1 & 3.92  & 48.7  & 60.0  & 48.2 \\
   15 & 13  & 2   & 1:4 & 3.63  & 46.8  & 55.6  & 45.0 \\
 20 & 5   & 15  & 2:1 & 6.74  & 46.2  & 63.1  & 46.1  \\ 
 20 & 15  & 5   & 2:1 & 4.72  & 50.6  & 63.7  & 50.2  \\
  20 & 15  & 5   & 1:4 & 4.48  & 50.2  & 62.0  & 49.1  \\	
 35 & 5   & 30  & 2:1 & 7.76  & 53.5  & 73.5 & 53.5   \\
 35 & 15  & 20  & 2:1 & 6.60  & 57.5  & 76.2  & 57.3  \\ 
  35 & 15  & 20  & 1:4 & 6.44  & 57.1  & 74.8  & 56.6  \\ 
 50 &  5  & 45  & 2:1 & 8.33  & 59.5  & 81.0  & 59.3   \\ 
 50 & 15  & 35  & 2:1 & 7.50  & 63.3 & 84.3  & 63.2  \\
  50 & 15  & 35  & 1:4 & 7.37  & 62.8 & 83.0  & 62.5  \\
\enddata
\tablecomments{Results of thermal evolution models for planets at 4.5
Gyr.}
\end{deluxetable}

\section{Atmosphere and Spectra} With these upper limits on the flux
from the planet's interior, we can begin to assess the state of the
planet's atmosphere.  To model the pressure--temperature (\emph{P--T})
profile, chemical abundances, and spectrum of the planet we use a model
atmosphere code with a long heritage, including Titan \cp{Mckay89},
Uranus \cp{MM99}, and also a wide array of brown dwarfs
\cp{Marley96,Morley14} and exoplanets \cp{Fortney08a,Marley12}.  The
predicted chemical equilibrium abundances follow the methods of
\ct{Visscher10}.

For the initial exploration of this paper we choose five values for the
intrinsic flux, \tint, of 20, 30, 40, 50, and 60 K.  There are also some
additional modeling choices and constraints, which have caveats.  At
these cold temperatures, methane (CH$_4$) condenses out and leaves the
gas phase.  In our normal modeling efforts of relatively warmer objects,
for instance water condensation into clouds in cool brown dwarfs, we
calculate the vapor pressure of water above any cloud, to
self-consistently determine the gaseous water mixing ratio at any
\emph{P--T} point \cp[e.g.,][]{Morley14}. Given the large uncertainties
for this possible planet, here we take a simpler path.  We choose an
opacity grid calculated at solar metallicity, but deplete these
opacities by various factors to simulate the loss of methane.   We also
use an opacity database that is only complete above 75 K for most
molecules (and 100 K for H$_2$) \cp{Freedman14}, such that the details
of the calculated spectra should be taken with some caution.  However,
our main findings below are insensitive to these caveats.

At a fiducial distance of 622 AU (motivated by Fienga et al.~2016), we present atmospheric \emph{P--T}
profiles in Figure 1a.  The most important conclusion from these models
is that for a wide range of parameter space the planet is cold enough
that most gaseous methane has condensed out of the planet's atmosphere,
leaving behind a nearly pure H/He atmosphere above a methane cloud that
could exist in a location between a few tenths of a bar to hundreds of
bars.  The vapor pressure (and hence mole fraction) of methane above the
cloud is \emph{independent of metallicity} and is directly tied to the
atmospheric temperature structure, such that a determination of the
methane mole fraction would serve as a ``thermometer" of the visible
atmosphere.

\begin{figure}[b!] \centering \epsscale{1.2} \plotone{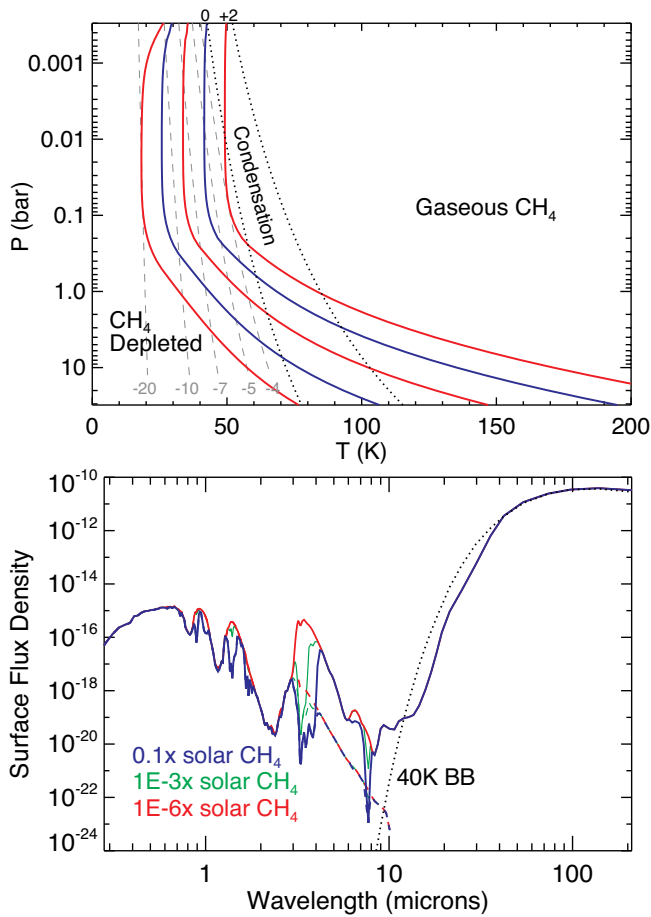}
\caption{\emph{Top:} Model \emph{P--T} profiles for Planet Nine,
assuming solar flux at 622 AU, with methane (CH$_4$) depleted to
0.1$\times$ the solar abundance.  The five profiles assume \tint=60, 50, 40,
30, and 20 K, right to left, and do not include cloud opacity.  In dotted black curves are condensation
curves for two abundances of methane: 100$\times$ solar (+2) and solar
(0).  To the right, methane is found in the gas phase, while to the
left, the mixing ratio of gaseous methane is severely depleted,
following the iso-abundance curves in gray, showing gaseous CH$_4$
mixing ratio.  A cloud of condensed methane will form in this
intermediate region. \emph{Bottom:} Model spectra (erg s$^{-1}$
cm$^{-2}$ Hz$^{-1}$, at the top of the atmosphere) from the optical to far infrared for three
\tint$=40$ K nominal models at 622 AU and with thick atmospheres and no
surface.  The dramatic change in spectra due to methane condensation is
shown, as methane is depleted at 0.1, 10$^{-3}$, and 10$^{-6}\times$
solar.  In the most depleted model, all absorption features are due to
H$_2$ CIA opacity.  The scattered light component is shown as dashed
curves.  A 40 K blackbody is shown as the black dotted curve.  Note the
tremendous enhancement in thermal fluxes above blackbody values at
$\sim$3-5 $\mu$m.} \label{profiles} \end{figure}

The spectra of these cold atmospheres with and without methane opacity
are shown in Figure 1b for a \tint$=40$ K model, representing the
expected upper-limit of the flux from a 10 \me model.  The optical
spectrum is nearly feature-free, except for H$_2$ CIA opacity at 0.8
$\mu$m and possibly methane, if present.  The profiles differ by a
factor of $10^5$ in the methane abundance, which is allowed given the
wide-range potential \emph{P--T} profiles.  In the mid-infrared, H$_2$
CIA and methane are the dominant absorbers, and the presence or absence
of methane dramatically alters the spectra.  In particular the thermal
fluxes from $\sim3-5$ $\mu$m are enhanced by around \emph{20 orders of
magnitude} compared to blackbody expectations.  All absorption features
in the red model spectrum are due to H$_2$ CIA.  In addition, the loss
of methane opacity shifts a prominent flux maximum from 4+ $\mu$m to
$\sim 3-3.5$ $\mu$m, which dramatically alters the flux ratio in the two
shortest \emph{WISE} and \emph{Spitzer} bands.  Tabulated thermal infrared
magnitudes are given in Table 2.

Our previously mentioned uncertain opacities at low temperature lead to some qualifications on our conclusions.  A strong enhancement of flux from 3-5$\mu$m is robust.  The large H$_2$ CIA opacities at tens of microns, with a pronounced minimum at 3-4 $\mu$m, is a generic outcome at any temperature \cp[e.g.][Figure 15]{Sharp07}, which forces flux into these opacities windows, as is well understood in brown dwarf and giant planet atmospheres.  However, quantitatively, the shape of the emitted spectra certainly do depend on specifics of the absorption coefficients.  An addition, we do not include the effects of the hypothesized, and potentially optically thick (at least at short wavelengths) CH$_4$ cloud, given the large uncertainties on its location and optical properties.  This could depress the 3-5 $\mu$m flux compared to the cloud-free case, which emerges from deep, hotter atmospheric layers.   Even with  potentially dramatic flux enhancements, given the faintness of the
planet in the mid-infared, we expect that in the near future a detection
in the optical is the most likely, and we next turn to what information
may be gleaned from an optical spectrum.

%
\begin{deluxetable*}{ccccccccccccccccccccccc}[b!] \tabletypesize{\scriptsize}
\tablecolumns{12} \tablewidth{0pt} \tablehead{ \colhead{$T_{\rm eff}$} &
\colhead{CH$_4$} & \colhead{u} & \colhead{g} & \colhead{r} & \colhead{i} & \colhead{z}& \colhead{Y} & \colhead{VR} & \colhead{Y} & \colhead{J} & \colhead{H} & \colhead{K} & \colhead{L'} & \colhead{M'} & \colhead{W1} & \colhead{W2}
& \colhead{W3} & \colhead{W4} \\
\colhead{(K)} & \colhead{abund}}

\startdata
  
20 & 10$^{-1}$ &  22.3 & 22.0 & 21.4 & 22.4 & 22.1 & 23.0 & 21.5 & 23.4 & 24.4 & 23.9 & 31.0 & 29.4 & 29.9 & 27.6 & 29.3 & 35.5 & 26.4\\
20 & 10$^{-3}$ &22.3&	22.0&		21.3&		22.3&		21.3&		21.9&		21.5&   22.4&	23.9&	23.4&	30.9&	28.1&	29.9&	27.1&	29.3&	36.4&	29.6\\
20 & 10$^{-6}$  & 22.3 & 22.0 & 21.3 & 22.3 & 21.2 & 21.8 & 21.5 & 22.3 & 23.8 & 23.4 & 30.9 & 27.1 & 29.9 & 26.0 & 29.3 & 36 .3 & 29.6\\
30 & 10$^{-1}$  & 22.3 & 22.0 & 21.4 & 22.4 & 22.2 & 23.0 & 21.5 & 23.3 & 24.3 & 24.0 & 30.8 & 29.5 & 29.7 & 27.4 & 29.0 & 31.9 & 19.8\\
30 & 10$^{-3}$  & 22.3 & 22.0 & 21.3 & 22.2 & 21.3 & 21.8 & 21.4 & 22.3 & 23.8 & 23.5 & 30.7 & 27.9 & 29.7 & 26.9 & 28.9 & 31.9 & 19.8\\
30 & 10$^{-6}$  & 22.3 & 22.0 & 21.3 & 22.2 & 21.2 & 21.8 & 21.4 & 22.2 & 23.7 & 23.5 & 30.7 & 26.4 & 29.7 & 25.6 & 28.9 & 31.9 & 19.8\\
40 & 10$^{-1}$  & 22.3 & 22.0 & 21.4 & 22.3 & 22.2 & 22.9 & 21.5  & 23.3 & 24.3 & 24.1 & 30.6 & 26.0 & 23.9 & 27.6 & 23.2 & 25.4 & 14.2\\
40 & 10$^{-3}$  & 22.3 & 22.0 & 21.3 & 22.2 & 21.4 & 21.9 & 21.4  & 22.3 & 23.9 & 23.7 & 30.6 & 22.4 & 23.9 & 23.9 & 22.8 & 25.4 & 14.2\\
40 & 10$^{-6}$  & 22.3 & 22.0 & 21.3 & 22.2 & 21.3 & 21.8 & 21.4 &  22.3 & 23.8 & 23.6 & 30.6 & 20.5 & 23.9 & 20.5 & 22.8 & 25.4 & 14.2\\
50 & 10$^{-1}$ &  22.3 & 22.0 & 21.4 & 22.3 & 22.3 & 23.0 & 21.5  & 23.3 & 24.4 & 24.5 & 30.5 & 21.1 & 17.7 & 28.0 & 17.6 & 20.0 & 10.6\\
50 & 10$^{-3}$ &  22.3 & 22.0 & 21.3 & 22.1 & 21.5 & 22.0 & 21.4 &  22.4 & 24.0 & 23.9 & 30.5 & 17.3 & 17.7 & 19.0 & 17.2 & 20.0 & 10.6\\
50 & 10$^{-6}$ &  22.3 & 22.0 & 21.3 & 22.1 & 21.4 & 21.9 & 21.4  & 22.3 & 24.0 & 23.9 & 30.5 & 15.8 & 17.7 & 16.1 & 17.2 & 20.0 & 10.6\\
60 & 10$^{-1}$ &  22.3 & 22.0 & 21.4 & 22.2 & 22.5 & 23.1 & 21.5 &  23.4 & 24.6 & 24.9 & 30.4 & 18.0 & 13.5 & 27.5 & 13.7 & 15.9 & 8.1\\
  
\enddata
\tablecomments{Apparent magnitudes at 622 AU \cp[see][]{Fienga2016},
using the described cloud-free atmosphere models and the radius of Neptune.  The first seven filters are for DECcam (u to VR), the next five or MKO filters, while the last four are for WISE (W1 to W4).  L' band and redder are dominated by thermal emission, which is dominated by intrinsic flux and therefore does not depend on the orbital separation, while shorter wavelength bands are dominated by solar reflection.  Therefore these thermal vs.~reflected magnitudes will change as a function of orbital distance in different ways.}
\end{deluxetable*}

\section{Optical Characterization} 
\begin{figure}[t!] \epsscale{1.2} \centering \plotone{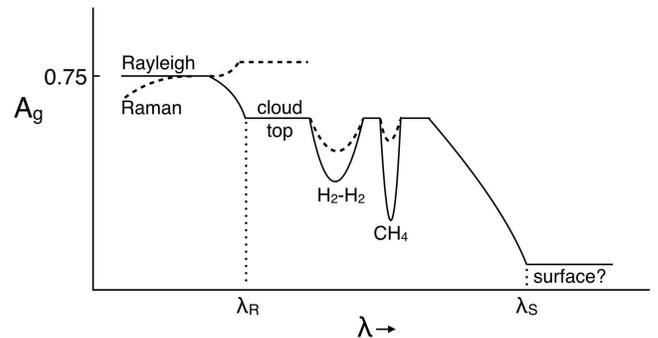}
\caption{Notional geometric albedo spectrum across the optical to near
infrared illustrates the expected key influences on the reflected light
spectrum. For a deep H$_2$ atmosphere most atmospheric opacity is due to
Rayleigh scattering, but since the cross-section falls off as
$\lambda^{-4}$, one sees progressively deeper at redder wavelengths. At
some wavelength $\lambda_{\rm R}$ one sees photons that reflect off of
the top of the CH$_4$ cloud, which could be brighter (dashed) or darker
than a purely Rayleigh scattering atmosphere. The band depths due to
H$_2$ and CH$_4$ depend, as illustrated by the dashed lines, on the
column density above the cloud and the gravity. At even longer
wavelengths, Mie cloud opacity lessens, and there could be reflection
from a planetary surface beyond $\lambda_{\rm S}$. At the bluest
wavelengths Raman scattering by H$_2$ molecules may be important, reducing
the geometric albedo below the pure Rayleigh limit of 0.75. }
\label{albedo} \end{figure}

Because any atmosphere of Planet Nine
will be cold and may be relatively ``clean" compared to solar system
analogs, the reflection spectrum may well have some counter-intuitive
characteristics, which we discuss below. In Figure \ref{albedo} we
present a notional geometric albedo spectrum of a planet such as modeled
here to help motivate early characterization efforts.

The geometric albedo of an infinitely deep Rayleigh scattering albedo is
0.75 at all wavelengths. \cp[See][for a discussion of the various forms
of albedo and limiting cases.]{Marley99} Thus the scattered light
spectrum from a deep, mostly $\rm H_2 - He$ atmosphere would primarily
be a mirror of that of the sun. At UV wavelengths, however, Raman
scattering by $\rm H_2$ becomes important \cp[e.g.,][]{Marley99,Oklopcic16} and will depress the geometric
albedo (to as low as $\sim 0.6$). But at longer wavelengths where the
Rayleigh cross section is high and Raman less important, $\sigma_{\rm
Ray}\approx{\sigma_0} \lambda^{-4}$ (where $\sigma_0=
8.14\times10^{-17}$ with $\lambda$ in nm), is large and the
photochemical hazes that darken the UV and blue spectra of solar system
giants will likely be sparse or absent (owing to the faint UV flux and
the paucity of methane). If there is, however, a bright $\rm CH_4$ cloud
deck then we would expect the geometric albedo to transition from the
pure Rayleigh value in the blue to a value controlled by Mie scattering
from the cloud deck at a wavelength $\lambda_{\rm R}$ where the column
two-way Rayleigh scattering optical depth, averaged over the disk, is
about 1: $2 \tau_{\rm Ray}/{\bar \mu} \approx 1$. Assuming hydrostatic
equilibrium in an approximately isothermal atmosphere and setting $\bar
\mu=0.5$ we have
\begin{equation}
\lambda_{\rm R}=\Bigl({4\sigma_0 {P_c}\over mg}\Bigr)^{1/4},
\end{equation}

where $m$ is the mean molecular weight of the atmosphere and
$g$ is the gravity, yielding $\lambda_{\rm R}$ in nm. For a cloudtop at
10 bar and $g=800\,\rm cm^2/s)$, $\lambda_{\rm R}=400\,\rm nm$. Whether
the cloud is brighter or darker than the overlying Rayleigh atmosphere
depends on its column mass, particle size, and vertical abundance
profile. Thus Figure \ref{albedo} illustrates both a brighter and darker
cloud influencing the spectra at $\lambda > \lambda_{\rm R}$.
Regardless, a break in the slope of the scattered light spectrum thus
provides a measure of the cloud top altitude, which helps to constrain
the atmosphere's thermal profile (see Figure 1a).

%
%
We also expect (Figure 1b) that there may well be some absorption
features evident in a reflected light spectrum of the planet, including
pressure-induced absorption features from $\rm H_2-H_2$ and $\rm H_2-He$
at $\sim~0.8\,\rm \mu m$.  The depth of the features will depend upon
the column of gas above the cloud. For deep, cloudless atmospheres these
features could be quite deep and broad. Finally gaseous CH$_4$
absorption above the cloud deck will produce the usual ``giant planet''
methane bands, although given the low temperatures and the expected
small column above the cloud it is may that in the optical only the strongest band, at
$0.889\,\rm \mu m$ will be prominent.  The colder the atmosphere, the
deeper the cloud, the redder $\lambda_{\rm R}$, and the lower the mixing
ratio of gaseous methane in the observable atmosphere.  The use of the CH$_4$ mixing ratio as a ``atmospheric thermometer" could be complicated since the mixing ratio will be a strong function of temperature (and hence, depth) in the atmosphere (Figure 1a).  This could necessitate the detection multiple bands of CH$_4$ (into the near IR).  These bands would be of different inherent strengths, which would become optically thick at different atmospheric depths, which could yield a full picture of the CH$_4$ mixing ratio with depth.
%

If there is a solid surface, then at some wavelength, $\lambda_{\rm S}$,
the scattered light albedo will be dominated by the properties of the
surface. For Mie scattering particles the extinction cross section drops
precipitously for wavelengths greater than about 10 times the mean
particle radius. Thus for a cloud consisting of submicron particles,
near-infrared flux may penetrate both the cloud and the scattering gas,
assuming the surface pressure is not too great. The windows through
which the deep atmosphere of Venus and the surface of Titan are glimpsed
through haze and gas are examples of this effect. The wavelength of such
a break at which the surface becomes detectable, if observed, provides a
measure of the surface pressure $\lambda_{\rm S} \propto P_{\rm
s}^{1/4}$.

%
%
%
\section{Detectability}

With its expected eccentricity of order $e=0.6$, and a posited
semi-major axis of order $a=700\,{\rm AU}$, Planet Nine's reflected
light visual magnitude would vary by a factor of $((1+e)/(1-e))^{4}=256$
over the course of its orbit. As a consequence, its detectability is
strongly influenced by its current radial distance from the Sun, which,
as a function of true anomaly, $\nu$, is given by
\begin{equation}
r_{P9}=\frac{a(1-e^2)}{(1+e\cos \nu)}\, .
\end{equation}

Planet Nine's V-band magnitude, therefore, is
\begin{equation}
V_{P9}=7.8+5\log_{10}\left[\left(\frac{R_{\rm P9}}{R_{\rm
Nep}}\right)^{2}\left(\frac{A_{\rm P9}}{A_{\rm
Nep}}\right)\left(\frac{r_{P9}}{29\,{\rm AU}}\right)^{4}\right]\, ,
\end{equation}
where $V_{\rm N}=7.8$ is Neptune's visual magnitude at opposition, and
Neptune's geometric albedo is taken as $A_{\rm Nep}=0.41$.

As a concrete illustration, consider the fiducial 10~$M_{\oplus}$ Planet
Nine candidate proposed by \citet{BatyginBrown2016}, in conjunction with
the orbital location obtained by \citet{Fienga2016} that provides the
maximum reduction of the Cassini residuals. These assumptions lead to
orbital elements: $a=700\,{\rm AU}$, $e=0.6$, $i=30^{\circ}$,
$\omega=150^{\circ}$, $\Omega=113^{\circ}$, and $\nu=118^{\circ}$, where
$i$ is the inclination to the ecliptic, $\omega$ is the argument of
perihelion, and $\Omega$ is the longitude of the planet's ascending
node. These elements correspond to a sky position of R.A.$\,\sim 2\,{\rm
hr}$, Dec$\,\sim-20^{\circ}$ in the constellation Cetus, and imply a
current distance $r_{\rm P9}=622\,{\rm AU}$.\footnote{Cetus is optimally
viewed in August through October in the Southern Hemisphere.} In
addition, if we adopt a Neptune-like envelope mass fraction of $f_{\rm
env}=M_{\rm atm}/M_{\rm pl}=0.1$, our structural model suggests $R_{\rm
P9}=3.46\,R_{\oplus}$. Drawing on the results of \S4, we adopt a
geometric albedo $A_{\rm P9}=0.75$, which generates a current visual
magnitude V=21.6. In the event that it has these parameters, Planet Nine
would be detectable in the optical by the ongoing Dark Energy Survey,
which is sensitive in the relevant sky location to $r<23.8$
\citep{BrownBatygin2016}.

\begin{figure}[h!] \epsscale{1.3} \centering
\plotone{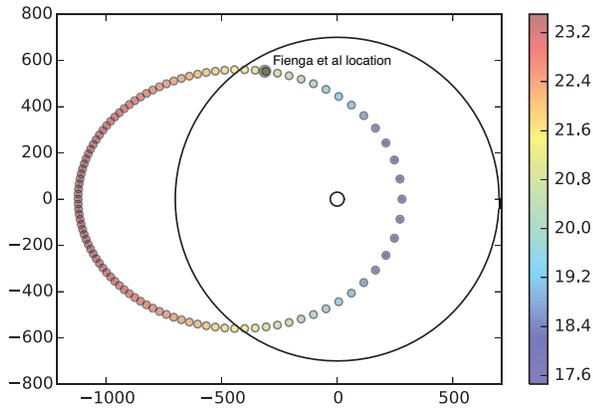} \caption{Planet Nine orbit with the
position of the planet shown at 100 equally spaced time intervals. Color
is tied to expected V magnitude, as computed with the atmospheric models
of the previous section. The 700 AU semi-major axis is indicated by the
black solid line. The best-fit location derived by \citet{Fienga2016} is
shown as a black dot.} \label{orbit} \end{figure}

If considered as a 40K blackbody radiator, Planet Nine's thermal
spectrum peaks at $\lambda\sim70\, \mu{\rm m}$, rendering its detection
difficult, but potentially possible, with present-day space-based
instrumentation and survey coverage.

\citet{Ginzburg2016} adopt a primarily analytic approach to infer the
current planetary surface temperature from the mass and composition of
the planet. Using a two-component  model for the core-envelope structure
of the planet, they show that some of the mass-composition parameter
space (assuming a black body spectrum) is accessible to the NASA WISE
Mission's W4 (22~$\mu$m) bandpass, and find that for a bulk composition
with $f_{\rm env}=M_{\rm atm}/M_{\rm pl}=0.1$, and $R_{\rm
pl}=1.2~R_{\rm Nep}$, Planet Nine is potentially detectable by WISE at
an $r_{P9}\sim a \sim$700 AU distance.

The models delineated in Table 2 indicate that Planet Nine might be much
brighter in the W1 (3.4~$\mu$m) WISE band than its effective temperature
would otherwise suggest. Our extremal case, with $T_{\rm eff}=50~{\rm
K}$ and $10^{-6}\,{\rm CH_{4}}$ abundance, predicts W1=16.1, which
exceeds the WISE All-Sky 5$\sigma$ sensitivity limit of W1=16.5
\citep{Wright10} for static object flux measurements obtained by
co-adding individual source frame images. For W1 and W2, these limits
were substantially improved (for stationary sources) through the
incorporation of post-cryo NEOWISE observations as well as by data
processing improvements \citep{Mainzer11}. On the other hand, an object
that moves significantly between epochs would contribute all of its flux
to a given catalogued point source during a single epoch. For example,
if P9 with intrinsic W1=16.11 were to contribute to half of the frames
in a co-added two-epoch image, its reported W1 would be $\sim$0.75
magnitudes fainter than its true W1.

At the nominal \citet{Fienga2016} location of R.A.$\,\sim 2\,{\rm hr}$,
Dec$\,\sim-20^{\circ}$, Planet Nine's total annual parallax shift would
be of order $\pi = 663{''}$, and its yearly positional shift on the sky
arising from its orbital motion would be of order $\sigma \sim
60{''}\,{\rm yr^{-1}}$. These displacements combine to yield a maximum
sky motion of $\delta \lesssim 6{''}\,{\rm day^{-1}}$. Given WISE's
sun-synchronous geocentric orbit and Planet Nine's current nominal
angular parameters ($i=30^{\circ}$, $\Omega=113^{\circ}$, and
$\nu=118^{\circ}$), its expected angular sky motion at the epochs of
observation would be $\dot{\pi}\sim 0.12''{\,\rm hr}^{-1}$, or $\Delta
\pi=2.3''$ over the 4/5th day observing sequence that is expected at
Planet Nine's current ecliptic latitude. This estimate for $\Delta \pi$
is smaller than WISE's W1 FWHM resolution of 6.1${''}$. As a
consequence, any WISE detection of Planet Nine in the AllWISE point
source
catalog\footnote{\url{http://wise2.ipac.caltech.edu/docs/release/allwise
/expsup/sec1\_5.html}} should register as a series of several fully
separated W1-detected (and possibly W2-detected) objects lying within a
roughly one square arc minute region of the sky. These separate
appearances of Planet Nine, in turn, will each be drawn from roughly
twelve  single-exposure images taken during a $\sim$4/5 day period
covered by overlapping scans; for further details see \citet{Wright10}.

We filtered the AllWISE Source Catalog for objects with $0~{\rm h}<{\rm
RA}<4~{\rm h}$ and $-30^{\circ}<{\rm Dec}<0^{\circ}$, having W1$>$16,
W2$>$17 (or null), W3$>20$ (or null), and W4$>$10 (or null). This screen
produced a single match: J005009.06 -315732.2 (J005), which has
W1=16.972 and W2=17.032, and which was observed by WISE on 5 individual
image frames. The WISE catalog reports that J005's W1 timestamps span a
single 2-day period from HJD 2455542.53 through 2455546.16. Based on the
foregoing discussion, a single-source detection built up over 4 days is
at best marginally consistent with Planet Nine's expected sky movement.
The J005 object is not, furthermore, paired with any sources having
similar photometry in the near vicinity and at separate epochs.
Moreover, J005's W2 single-frame detection epochs range over a 6-month
span from HJD 2455365.12 to HJD 2455546.16, implying that it is a
transiently detected, yet nonetheless stationary object or noise source.
J005 can therefore confidently be rejected from consideration as a
Planet Nine candidate. While we believe it is exceedingly unlikely that
Planet Nine can be located in the AllWISE Source Catalog, it is unlikely
yet nonetheless possible that it makes an appearance in the AllWISE
Reject Table. We are currently sifting this 428,787,253-line database
for candidates.

\section{Conclusions} We have modeled several aspects of the thermal
evolution and atmospheric structure and spectra of the possible Planet
Nine.  Like others, we find a planetary \teff\ of perhaps $35-50$ K,
depending on mass, but we caution that these are upper limits.  The
atmosphere could well be in a temperature range of significant
condensation of methane.  The loss of this important ``giant planet''
opacity source can dramatically alter both the reflection and emission
spectra away from expectations based on Uranus and Neptune, with the
notable effect of causing Planet Nine to appear ``blue" rather than
``red" in the WISE bands.

Our models suggest that the reflection spectrum, due to a mix of
Rayleigh scattered sunlight, cloud scattering, and absorption due to
H$_2$ and CH$_4$ opacity could provide rich diagnostics of the
temperature structure, possible cloud location, and even surface
pressure of the atmosphere.  Let the hunt continue.

\acknowledgements
We thank the anonymous referees for helpful comments on the draft, Johanna Teske and Antonija Oklopicic for wise insights, and Kevin Luhman for WISE insights.

\end{document}